\documentclass[a4paper,11pt]{article}

\usepackage{jheppub} 

\title{\boldmath Evolution equation for 3-quark Wilson loop operator}

 \author{R. E. Gerasimov}
 \author{and A. V. Grabovsky}
 \affiliation{Budker Institute of Nuclear Physics and Novosibirsk State University,\\630090 Novosibirsk, Russia}

\emailAdd{R.E.Gerasimov@inp.nsk.su}
\emailAdd{A.V.Grabovsky@inp.nsk.su}

\abstract{The evolution equation for the 3 quark Wilson loop operator has been derived in the leading logarithm approximation within Balitsky high energy operator expansion.}

\arxivnumber{}

\begin{document} 
\maketitle
\flushbottom

\section{Introduction}

There are two principal approaches to the description of semihard processes in
QCD. The first one is based on the Balitsky-Fadin-Kuraev-Lipatov (BFKL)
equation \cite{Fadin:1975cb}--\cite{Balitsky:1978ic}, while the second one stems from the dipole picture of
high energy scattering \cite{Nikolaev:1993th}--\cite{Mueller:1994jq} and is based on the Balitsky-Kovchegov
(BK) equation \cite{Balitsky}--\cite{Kovchegov}. Both these approaches are now
developed in the next to leading approximation (see \cite{Fadin:1998py}--\cite{Fadin:2007xy} and
\cite{Balitsky:2006wa}--\cite{Balitsky:2009xg}, respectively) and it was shown that for
the scattering of colorless particles in the linear regime they coincide
\cite{Fadin:2006ha}--\cite{Fadin:2009gh}.

In many technical details of these approaches special attention was payed to
the specific form of the colliding particle, namely to the color dipole. At the same time there is one more fundamental object. It is the colorless color triplet Green function, describing the evolution of the
proton Green function with energy. This case was analyzed first in
\cite{Praszalowicz:1997nf}, where the linear evolution was studied. Later in
\cite{Bartels:2007aa} the nonlinear evolution equation and impact factors for
baryon scattering were obtained. Here this case is studied in the leading
logarithm approximation within the Wilson line approach, proposed by I. Balitsky \cite{Balitsky},\cite{Balitsky:2010jf}%
. In fact in this paper, the same steps are done to come to the evolution equation  for the 3-quark Wilson line
$\varepsilon^{i^{\prime}j^{\prime}h^{\prime}}\varepsilon_{ijh}U_{1i^{\prime}%
}^{i}U_{2j^{\prime}}^{j}U_{3h^{\prime}}^{h},$ as in \cite{Balitsky},\cite{Balitsky:2010jf} for the
color dipole operator.

The paper is organized as follows. In the next section the leading order
evolution equation for the 3-quark Wilson loop is derived. In section 3 and 4
the cases of C-even and C-odd exchanges are discussed. Section 5 concludes
the paper.

\section{3-quark Wilson loop evolution equation}

We introduce the light cone vectors $n_{1}$ and $n_{2}$%
\begin{equation}
n_{1}=\left(  1,0,0,1\right)  ,\quad n_{2}=\frac{1}{2}\left(  1,0,0,-1\right)
,\quad n_{1}^{+}=n_{2}^{-}=n_{1}n_{2}=1
\end{equation}
and for any vector $p$ we have%
\begin{equation}
p^{+}=p_{-}=pn_{2}=\frac{1}{2}\left(  p^{0}+p^{3}\right)  ,\qquad p_{+}%
=p^{-}=pn_{1}=p^{0}-p^{3},
\end{equation}%
\begin{equation}
p=p^{+}n_{1}+p^{-}n_{2}+p_{\bot},\qquad p^{2}=2p^{+}p^{-}-\vec{p}%
^{\,2},
\end{equation}%
\begin{equation}
\quad p\,k=p^{\mu}k_{\mu}=p^{+}k^{-}+p^{-}k^{+}-\vec{p}\vec{k}=p_{+}%
k_{-}+p_{-}k_{+}-\vec{p}\vec{k}.
\end{equation}
We would like to write an evolution equation for the 3-quark Wilson loop
operator%
\begin{equation}
B_{123}^{\eta}=\varepsilon^{i^{\prime}j^{\prime}h^{\prime}}\varepsilon
_{ijh}U\left(  \vec{z}_{1},\eta\right)  _{i^{\prime}}^{i}U\left(  \vec{z}%
_{2},\eta\right)  _{j^{\prime}}^{j}U\left(  \vec{z}_{3},\eta\right)
_{h^{\prime}}^{h} \label{B}%
\end{equation}
describing baryon scattering off the shock wave. Hereafter we will use the
following notation for such convolutions%
\begin{equation}
\varepsilon^{i^{\prime}j^{\prime}h^{\prime}}\varepsilon_{ijh}U_{1i^{\prime}%
}^{i}U_{2j^{\prime}}^{j}U_{3h^{\prime}}^{h}=U_{1}\cdot U_{2}\cdot U_{3}.
\end{equation}
Here%
\begin{equation}
U\left(  \vec{z},\eta\right)  =P e^{ig\int_{-\infty}^{+\infty}b_{\eta}%
^{-}(z^{+},\vec{z})dz^{+}}, \label{WL}%
\end{equation}
and $b_{\eta}^{-}$ is the external shock wave field built from only slow
gluons
\begin{equation}
b_{\eta}^{-}=\int\frac{d^{4}p}{\left(  2\pi\right)  ^{4}}e^{-ipz}b^{-}\left(
p\right)  \theta(e^{\eta}-p^{+}).
\end{equation}
The parameter $\eta$ separates the slow gluons entering the Wilson lines from
the fast ones in the impact factors. The shape of the path at $z^{+}=\pm
\infty$ in (\ref{WL}) is not important because the field is concentrated at
$z^{+}=0$%
\begin{equation}
b^{\mu}\left(  z\right)  =b^{-}(z^{+},\vec{z})n_{2}^{\mu}=\delta
(z^{+})b\left(  \vec{z}\right)  n_{2}^{\mu}.
\end{equation}
Therefore one can connect the three lines in (\ref{B}) in one point $x$ at
$z^{+}=+\infty$ and in one point $y$ at $z^{+}=-\infty.$ Then the operator
$B_{123}^{\eta}$ is evidently gauge invariant since under a gauge rotation the
Wilson lines change
\begin{equation}
U\left(  \vec{z}_{1},\eta\right)  _{i^{\prime}}^{i}\rightarrow V(x)_{k}^{i}U\left(  \vec{z}_{1},\eta\right)  _{k^{\prime}}^{k}V(y)_{i^{\prime}%
}^{k^{\prime}},\quad V\in SU(3).
\end{equation}
The gluon propagator in the shock wave background reads%
\[
G_{\mu\nu}^{\eta}(x,y)^{ab}|_{x^{+}>0>y^{+}}=-\int\frac{dp^{+}}{\left(
2\pi\right)  ^{3}}\frac{p^{+}\theta\left(  p^{+}\right)  }{2x^{+}y^{+}}\int
d\vec{z}e^{-ip^{+}\left\{  x^{-}-y^{-}+\frac{(\vec{z}-\vec{y})^{2}+i0}{2y^{+}%
}-\frac{(\vec{x}-\vec{z})^{2}+i0}{2x^{+}}\right\}  }%
\]%
\begin{equation}
\times\frac{g_{\bot\mu}^{\,\,\,\,\,\,\alpha}x^{+}-(x-z)_{\bot}^{\alpha}%
n_{2\mu}}{x^{+}}U^{ab}(\vec{z},\eta)\frac{g_{\bot\alpha\nu}(-y^{+})-(z-y)_{\bot
\alpha}n_{2\nu}}{-y^{+}}. \label{gluon_prop_through_exponontial}%
\end{equation}
To derive the evolution equation we have to change $\eta\rightarrow\eta
+\Delta\eta$
\begin{equation}
b_{\eta_{1}}^{-}=b_{\eta_{2}}^{-}+b_{\Delta\eta}^{-},\quad b_{\Delta\eta}%
^{-}(z^{+},\vec{z})=\int\frac{d^{4}p}{\left(  2\pi\right)  ^{4}}e^{-ipz}%
b^{-}\left(  p\right)  \theta(e^{\eta_{1}}-p^{+})\theta(p^{+}-e^{\eta_{2}}).
\end{equation}
Therefore we have to include the gluons with $\eta_{1}>\ln p^{+}>\eta_{2}$
into the Wilson lines, namely
\begin{equation}
B_{123}^{\eta_{1}}=B_{123}^{\eta_{2}}+\frac{\langle0|T(B_{123}^{\Delta\eta
}e^{i\int\mathcal{L}\left(  z\right)  dz})|0\rangle}{\langle0|T(e^{i\int
\mathcal{L}\left(  z\right)  dz})|0\rangle}.
\end{equation}
Via the Wilson line definition
\[
U(\vec{z},\eta)=...\left(  1+igb_{\eta}^{-}(z^{+}+\Delta z^{+},\vec{z})\Delta
z^{+}\right)  
\]
\begin{equation}\times\left(  1+igb_{\eta}^{-}(z^{+},\vec{z})\Delta z^{+}\right)
\left(  1+igb_{\eta}^{-}(z^{+}-\Delta z^{+},\vec{z})\Delta z^{+}\right)  ...,
\label{WL2}%
\end{equation}
we get%
\begin{equation}
B_{123}^{\eta_{1}}=B_{123}^{\eta_{2}}+\Delta B_{r}+\Delta B_{v}.
\end{equation}
One can find the real contribution $\Delta B_{r}$ using
(\ref{gluon_prop_through_exponontial}) and (\ref{WL2})
\[
\Delta B_{r}=\frac{\Delta\eta\alpha_{s}}{\pi^{2}}\int d\vec{z}_{4}U^{ab}%
(\vec{z}_{4},\eta_{2})
\]%
\[
\times\left[  \frac{1}{\vec{z}_{41}^{\,\,2}}\left(  t^{a}U\left(  \vec{z}%
_{1},\eta_{2}\right)  t^{b}\right)  \cdot U\left(  \vec{z}_{2},\eta
_{2}\right)  \cdot U\left(  \vec{z}_{3},\eta_{2}\right)  +(1\leftrightarrow
2)+(1\leftrightarrow3)\right.
\]%
\[
+\frac{\vec{z}_{41}\vec{z}_{42}}{\vec{z}_{41}^{\,\,2}\vec{z}_{42}^{\,\,2}%
}\left(  (  t^{a}U\left(  \vec{z}_{1},\eta_{2}\right)  )
\cdot(  U\left(  \vec{z}_{2},\eta_{2}\right)  t^{b})  +(
U\left(  \vec{z}_{1},\eta_{2}\right)  t^{b})  \cdot\left(  t^{a}U\left(
\vec{z}_{2},\eta_{2}\right)  \right)  \right)  \cdot U\left(  \vec{z}_{3}%
,\eta_{2}\right)
\]%
\begin{equation}
\left.  +(1\leftrightarrow3)+(2\leftrightarrow3)\frac{}{}\right]  .
\end{equation}
The virtual contribution is easier to find from the requirement that there is
no evolution in $\eta$ without the shock-wave. Hereafter we will write $N_c=3$ explicitly since the form of the totally antisymmetric tensor depends on $N_c$. Taking into account
\begin{equation}
\varepsilon^{ijh}\varepsilon_{ijh}=6,\quad t^{a}t^{a}=\frac{4}{3},\quad\varepsilon^{i^{\prime}j^{\prime}h^{\prime}}\varepsilon
_{ijh}(t^{a})_{i^{\prime}}^{i}\left(  t^{a}\right)  _{j^{\prime}}^{j}%
=-\frac{2}{3}\varepsilon^{i^{\prime}j^{\prime}h^{\prime}%
}\varepsilon_{ijh}\delta_{i^{\prime}}^{i}\delta_{j^{\prime}}^{j},
\end{equation}%
\begin{equation}
\varepsilon^{i^{\prime}j^{\prime}h^{\prime}}\varepsilon_{ijh}\left(
t^{a}U\left(  \vec{z}_{1},\eta_{2}\right)  \right)  _{i^{\prime}}^{i}\left(
t^{a}U\left(  \vec{z}_{2},\eta_{2}\right)  \right)  _{j^{\prime}}^{j}%
=-\frac{2}{3}\varepsilon^{i^{\prime}j^{\prime}h^{\prime}%
}\varepsilon_{ijh}U\left(  \vec{z}_{1},\eta_{2}\right)  _{i^{\prime}}%
^{i}U\left(  \vec{z}_{2},\eta_{2}\right)  _{j^{\prime}}^{j},
\end{equation}%
 we find%
\[
\Delta B_{v}=-\frac{\Delta\eta\alpha_{s}}{\pi^{2}}\frac{2}{3}\int
d\vec{z}_{4}%
\]%
\begin{equation}
\times\left[  \frac{\vec{z}_{12}^{\,\,2}}{\vec{z}_{41}^{\,\,2}\vec{z}%
_{42}^{\,\,2}}+\frac{\vec{z}_{13}^{\,\,2}}{\vec{z}_{41}^{\,\,2}\vec{z}%
_{43}^{\,\,2}}+\frac{\vec{z}_{23}^{\,\,2}}{\vec{z}_{42}^{\,\,2}\vec{z}%
_{43}^{\,\,2}}\right]  U\left(  \vec{z}_{1},\eta_{2}\right)  \cdot U\left(
\vec{z}_{2},\eta_{2}\right)  \cdot U\left(  \vec{z}_{3},\eta_{2}\right)  .
\end{equation}
Since
\begin{equation}
U^{ab}( \vec{z}_{4},\eta_{2})=2tr\left(  t^{a}U\left(  \vec
{z}_{4},\eta_{2}\right)  t^{b}U\left(  \vec{z}_{4},\eta_{2}\right)  ^{\dag
}\right)  ,
\end{equation}
we have
\[
U^{ab}(\vec{z}_{4},\eta_{2})\left(  t^{a}U\left(  \vec{z}_{1},\eta_{2}\right)
t^{b}\right)  _{i^{\prime}}^{i}
\]
\begin{equation}
=\frac{1}{2}tr\left(  U\left(  \vec{z}_{1}%
,\eta_{2}\right)  U\left(  \vec{z}_{4},\eta_{2}\right)  ^{\dag}\right)
U\left(  \vec{z}_{4},\eta_{2}\right)  _{i^{\prime}}^{i}-\frac{1}{6}U\left(  \vec{z}_{1},\eta_{2}\right)  _{i^{\prime}}^{i},
\end{equation}%
\[
U^{ab}(\vec{z}_{4},\eta_{2})\left(  t^{a}U\left(  \vec{z}_{1},\eta_{2}\right)
\right)  _{i^{\prime}}^{i}\left(  U\left(  \vec{z}_{2},\eta_{2}\right)
t^{b}\right)  _{j^{\prime}}^{j}%
\]%
\begin{equation}
=\frac{1}{2}\left(  U\left(  \vec{z}_{2},\eta_{2}\right)  U\left(  \vec{z}%
_{4},\eta_{2}\right)  ^{\dag}U\left(  \vec{z}_{1},\eta_{2}\right)  \right)
_{i^{\prime}}^{j}U\left(  \vec{z}_{4},\eta_{2}\right)  _{j^{\prime}}^{i}%
-\frac{1}{6}U\left(  \vec{z}_{1},\eta_{2}\right)  _{i^{\prime}}%
^{i}U\left(  \vec{z}_{2},\eta_{2}\right)  _{j^{\prime}}^{j}.
\end{equation}
Therefore the evolution equation reads%
\[
\frac{\partial B_{123}^{\eta}}{\partial\eta}=\frac{\alpha_{s}}{2\pi^{2}}\int
d\vec{z}_{4}\left[  \left\{  \frac{C_{1}}{\vec{z}_{41}^{\,\,2}}%
+(1\leftrightarrow2)+(1\leftrightarrow3)\right\}  \right.
\]%
\begin{equation}
+\left.  \left\{  \frac{\vec{z}_{41}\vec{z}_{42}}{\vec{z}_{41}^{\,\,2}\vec
{z}_{42}^{\,\,2}}C_{12}+(1\leftrightarrow3)+(2\leftrightarrow3)\right\}
\right]  . \label{evolutionEq1}%
\end{equation}
where%
\begin{equation}
C_{1}=tr\left(  U\left(  \vec{z}_{1},\eta\right)  U\left(  \vec{z}_{4}%
,\eta\right)  ^{\dag}\right)  B_{423}^{\eta}-3 B_{123}^{\eta},
\end{equation}%
\[
C_{12}=2B_{123}^{\eta}
\]
\begin{equation}
-\left(  U\left(  \vec{z}_{2},\eta\right)  U\left(
\vec{z}_{4},\eta\right)  ^{\dag}U\left(  \vec{z}_{1},\eta\right)  +U\left(
\vec{z}_{1},\eta\right)  U\left(  \vec{z}_{4},\eta\right)  ^{\dag}U\left(
\vec{z}_{2},\eta\right)  \right)  \cdot U\left(  \vec{z}_{4},\eta\right)
\cdot U\left(  \vec{z}_{3},\eta\right)  .
\end{equation}
For brevity we will simplify the notation henceforth%
\begin{equation}
U\left(  \vec{z}_{1},\eta\right)  \equiv U_{1}.
\end{equation}
Wilson lines are elements of SU(3). Hence, we have%
\begin{equation}
\varepsilon^{i^{\prime}j^{\prime}h^{\prime}}U_{i^{\prime}}^{i}U_{j^{\prime}%
}^{j}U_{h^{\prime}}^{h}=\varepsilon^{ijh},
\end{equation}
\begin{equation}
\varepsilon_{ijh}%
\varepsilon^{i^{\prime}j^{\prime}h^{\prime}}U_{i^{\prime}}^{i}U_{j^{\prime}%
}^{j}=2(U^{\dag})_{h}^{h^{\prime}},\quad\varepsilon_{ijh}\varepsilon
^{i^{\prime}j^{\prime}h^{\prime}}(U^{\dag})_{i^{\prime}}^{i}(U^{\dag
})_{j^{\prime}}^{j}=2U_{h}^{h^{\prime}}, \label{SU3identities}%
\end{equation}%
\begin{equation}
U_{i}\cdot U_{j}\cdot U_{k}=(U_{i}U_{l}^{\dag})\cdot(U_{j}U_{l}^{\dag}%
)\cdot(U_{k}U_{l}^{\dag}). \label{3qWlidentity}%
\end{equation}%
\begin{equation}
B_{iij}^{\eta}=U_{i}\cdot U_{i}\cdot U_{j}=2tr(U_{j}U_{i}^{\dag}).
\label{113identity}%
\end{equation}
In fact, the latter identity states that quark-diquark and quark-antiquark
systems are described by the same operator. As a result,%
\begin{equation}
C_{1}=\frac{1}{2}B_{144}^{\eta}B_{423}^{\eta}-3 B_{123}^{\eta};
\end{equation}%
\[
C_{12}=2B_{123}^{\eta}-\left(  U_{2}U_{4}^{\dag}U_{1}+U_{1}U_{4}^{\dag}%
U_{2}\right)  \cdot U_{4}\cdot U_{3}%
\]%
\begin{equation}
=3B_{123}^{\eta}-\frac{1}{2}(B_{144}^{\eta}B_{324}^{\eta}+B_{244}^{\eta
}B_{314}^{\eta}-B_{344}^{\eta}B_{214}^{\eta}). \label{IDENTITY}%
\end{equation}
This identity follows from (\ref{3qWlidentity}-\ref{113identity}) directly.
Indeed, one can rewrite $C_{12}$ as
\[
C_{12}=2(U_{1}U_{4}^{\dag})\cdot(U_{2}U_{4}^{\dag})\cdot(U_{3}U_{4}^{\dag
})-\left(  U_{2}U_{4}^{\dag}U_{1}U_{4}^{\dag}+U_{1}U_{4}^{\dag}U_{2}%
U_{4}^{\dag}\right)  \cdot E\cdot(U_{3}U_{4}^{\dag})
\]
and the right hand side of (\ref{IDENTITY}) as%
\[
\text{r.h.s. of (\ref{IDENTITY})}=3(U_{1}U_{4}^{\dag})\cdot(U_{2}U_{4}^{\dag
})\cdot(U_{3}U_{4}^{\dag})
\]%
\[
-tr(U_{1}U_{4}^{\dag})(U_{3}U_{4}^{\dag})\cdot(U_{2}U_{4}^{\dag})\cdot
E-tr(U_{2}U_{4}^{\dag})(U_{3}U_{4}^{\dag})\cdot(U_{1}U_{4}^{\dag})\cdot E
\]%
\begin{equation}
+tr(U_{3}U_{4}^{\dag})(U_{2}U_{4}^{\dag})\cdot(U_{1}U_{4}^{\dag})\cdot E.
\end{equation}
Next, expanding the Levi-Civita symbols as
\begin{equation}
\varepsilon_{ijh}\varepsilon^{i^{\prime}j^{\prime}h^{\prime}}=\left\vert
\begin{tabular}
[c]{lll}%
$\delta_{i}^{i^{\prime}}$ & $\delta_{i}^{j^{\prime}}$ & $\delta_{i}%
^{h^{\prime}}$\\
$\delta_{j}^{i^{\prime}}$ & $\delta_{j}^{j^{\prime}}$ & $\delta_{j}%
^{h^{\prime}}$\\
$\delta_{h}^{i^{\prime}}$ & $\delta_{h}^{j^{\prime}}$ & $\delta_{h}%
^{h^{\prime}}$%
\end{tabular}
\right\vert ,
\end{equation}
one proves (\ref{IDENTITY}). Finally, one gets the following evolution
equation%
\[
\frac{\partial B_{123}^{\eta}}{\partial\eta}=\frac{\alpha_{s}3}{4\pi^{2}%
}\int d\vec{z}_{4}\left[  \frac{\vec{z}_{12}^{\,\,2}}{\vec{z}_{41}^{\,\,2}%
\vec{z}_{42}^{\,\,2}}(-B_{123}^{\eta}+\frac{1}{6}(B_{144}^{\eta}%
B_{324}^{\eta}+B_{244}^{\eta}B_{314}^{\eta}-B_{344}^{\eta}B_{214}^{\eta
}))\right.
\]%
\begin{equation}
\left.\frac{}{}  +(1\leftrightarrow3)+(2\leftrightarrow3)\right]  .
\label{evolutionEq2}%
\end{equation}
Equation (\ref{evolutionEq2}) has several important properties. First, it has
no singularities at $\vec{z}_{4}=\vec{z}_{1,2,3}.$ Then, this equation changes
into the BK equation if two of the three quark coordinates coincide. One can
check it straightforwardly using (\ref{113identity}). It means that $q\bar{q}$
and $q$-diquark systems obey the same equation.

\section{C-even exchange}

To separate the C-odd and the C-even contributions we have to write down
the evolution equation for $B_{\bar{1}\bar{2}\bar{3}}^{\eta},$ i.e. the
3-antiquark Wilson loop operator%
\begin{equation}
B_{\bar{1}\bar{2}\bar{3}}^{\eta}=U_{1}^{\dag}\cdot U_{2}^{\dag}\cdot
U_{3}^{\dag},
\end{equation}
describing antibaryon scattering off the shock wave. One can get such an
equation from the equation for $B_{123}^{\eta}$ changing all the Wilson lines
to their conjugates $U_{i}\leftrightarrow U_{i}^{\dag}.$ The C-even Green
function has the following form%
\begin{equation}
B_{123}^{+}=B_{123}^{\eta}+B_{\bar{1}\bar{2}\bar{3}}^{\eta}-12%
.\label{pomeronFG}%
\end{equation}
The operator%
\begin{equation}
B_{123}^{-}=B_{123}^{\eta}-B_{\bar{1}\bar{2}\bar{3}}^{\eta}\label{odderonFG}%
\end{equation}
changes its sign under $C$ transformation, hence it describes the C-odd
Green function. Rewriting evolution equation (\ref{evolutionEq2}) in terms of
(\ref{pomeronFG}) and (\ref{odderonFG}), we have
\[
\frac{\partial B_{123}^{+}}{\partial\eta}=\frac{\alpha_{s}3}{4\pi^{2}}\int
d\vec{z}_{4}\frac{\vec{z}_{12}^{\,\,2}}{\vec{z}_{14}^{\,\,2}\vec{z}%
_{42}^{\,\,2}}\left[  B_{144}^{+}+B_{244}^{+}-B_{344}^{+}\frac{{}}{{}}\right.
\]%
\[
+B_{134}^{+}+B_{234}^{+}-B_{123}^{+}-B_{124}^{+}+\frac{1}{12}\left(
B_{144}^{+}B_{324}^{+}+B_{244}^{+}B_{314}^{+}-B_{344}^{+}B_{214}^{+}\right)
\]%
\begin{equation}
+\left.  \frac{1}{12}\left(  B_{144}^{-}B_{324}^{-}+B_{244}^{-}B_{314}%
^{-}-B_{344}^{-}B_{214}^{-}\right)  \right]  +(2\leftrightarrow
3)+(1\leftrightarrow3).
\end{equation}
Pomeron exchange starts from the 2-gluon exchange. One can show that
\begin{equation}
B_{123}^{+}=\frac{1}{2}(B_{133}^{+}+B_{211}^{+}+B_{322}^{+})+\tilde{B}%
_{123}^{+}.\label{B123plus}%
\end{equation}
where $\tilde{B}_{123}^{+}$\ works from the 4-gluon exchange. Indeed, in the
2- and 3-gluon approximations one can consider
\[
0=(U_{1}-U_{2})\cdot(U_{2}-U_{3})\cdot(U_{3}-U_{2})+(U_{1}^{\dag}-U_{2}^{\dag
})\cdot(U_{2}^{\dag}-U_{3}^{\dag})\cdot(U_{3}^{\dag}-U_{2}^{\dag})=
\]%
\begin{equation}
=2B_{123}^{+}-B_{122}^{+}-B_{133}^{+}-B_{223}^{+},
\end{equation}
which is clear in the 2-gluon approximation, and which follows from 
\[
(U_{1}-U_{2})\cdot(U_{2}-U_{3})\cdot(U_{3}-U_{2})
\]%
\[
=U_{1}^{\dag}(U_{1}-U_{2})U_{2}^{\dag}\cdot U_{2}^{\dag}(U_{2}-U_{3}%
)U_{3}^{\dag}\cdot U_{3}^{\dag}(U_{3}-U_{2})U_{2}^{\dag}%
\]%
\begin{equation}
=(U_{2}^{\dag}-U_{1}^{\dag})\cdot(U_{3}^{\dag}-U_{2}^{\dag})\cdot(U_{2}^{\dag
}-U_{3}^{\dag})
\end{equation}
in the 3-gluon one.
As a result, one can write the linear equation in the 2- and 3-gluon
approximations for $B_{123}^{+}$%
\[
\frac{\partial B_{123}^{+}}{\partial\eta}=\frac{1}{2}\frac{\partial}%
{\partial\eta}(B_{133}^{+}+B_{211}^{+}+B_{322}^{+})
\]%
\begin{equation}
=\frac{\alpha_{s}3}{4\pi^{2}}\int d\vec{z}_{4}\left[  \frac{\vec{z}%
_{12}^{\,\,2}}{\vec{z}_{14}^{\,\,2}\vec{z}_{42}^{\,\,2}}\left(  B_{144}%
^{+}+B_{244}^{+}-B_{122}^{+}\right)  +(2\leftrightarrow3)+(1\leftrightarrow
3)\right]  ,
\end{equation}
which is a sum of 3 independent BFKL equations.

In further approximations one should take into account $\tilde{B}_{ijk}^{+}.$
It leads us to the following evolution equation%
\[
\frac{\partial B_{123}^{+}}{\partial\eta}=\frac{1}{2}\frac{\partial}%
{\partial\eta}(B_{133}^{+}+B_{211}^{+}+B_{322}^{+})+\frac{\partial\tilde
{B}_{123}^{+}}{\partial\eta}%
\]%
\[
=\frac{\alpha_{s}3}{4\pi^{2}}\int d\vec{z}_{4}\frac{\vec{z}_{12}^{\,\,2}%
}{\vec{z}_{14}^{\,\,2}\vec{z}_{42}^{\,\,2}}\left[  \left(  B_{144}^{+}%
+B_{244}^{+}-B_{122}^{+}\right)  +(\tilde{B}_{134}^{+}+\tilde{B}_{234}%
^{+}-\tilde{B}_{123}^{+}-\tilde{B}_{124}^{+})\frac{{}}{{}}\right.
\]%
\[
+\frac{1}{24}\left(  B_{144}^{+}B_{322}^{+}+2B_{144}^{+}B_{224}%
^{+}+B_{244}^{+}B_{311}^{+}-B_{344}^{+}B_{211}^{+}\right)
\]%
\[
+\frac{1}{12}\left(  B_{144}^{+}\tilde{B}_{324}^{+}+B_{244}^{+}\tilde
{B}_{314}^{+}-B_{344}^{+}\tilde{B}_{214}^{+}\right)
\]%
\begin{equation}
+\left.  \frac{1}{12}\left(  B_{144}^{-}B_{324}^{-}+B_{244}^{-}B_{314}%
^{-}-B_{344}^{-}B_{214}^{-}\right)  \right]  +(2\leftrightarrow
3)+(1\leftrightarrow3). \label{eq4+}%
\end{equation}
The dipole BK equation has the form
\begin{equation}
\frac{\partial tr(U_{1}U_{2}^{\dag})}{\partial\eta}=\frac{\alpha_{s}}{2\pi
^{2}}\int d\vec{z}_{4}\frac{\vec{z}_{12}^{\,\,2}}{\vec{z}_{14}^{\,\,2}\vec
{z}_{42}^{\,\,2}}\left[  tr(U_{1}U_{4}^{\dag})tr(U_{4}U_{2}^{\dag}%
)-N_{c}tr(U_{1}U_{2}^{\dag})\right]  . \label{BK}%
\end{equation}
For C-even exchanges it can be rewritten in our notation as%
\begin{equation}
\frac{\partial}{\partial\eta}\frac{1}{2}B_{211}^{+}=\frac{\alpha_{s}N_{c}%
}{4\pi^{2}}\int d\vec{z}_{4}\frac{\vec{z}_{12}^{\,\,2}}{\vec{z}_{14}%
^{\,\,2}\vec{z}_{42}^{\,\,2}}\left[  \left(  B_{144}^{+}+B_{244}^{+}%
-B_{122}^{+}\right)  +\frac{1}{4N_{c}}\left(  B_{144}^{+}B_{422}^{+}%
+B_{144}^{-}B_{422}^{-}\right)  \right]  . \label{BKeven}%
\end{equation}
Taking $N_c=3$ and subtracting 3 BK equations from (\ref{eq4+}) one comes to%
\[
\frac{\partial\tilde{B}_{123}^{+}}{\partial\eta}=\frac{\alpha_{s}3}%
{4\pi^{2}}\int d\vec{z}_{4}\frac{\vec{z}_{12}^{\,\,2}}{\vec{z}_{14}%
^{\,\,2}\vec{z}_{42}^{\,\,2}}\left[  (\tilde{B}_{134}^{+}+\tilde{B}_{234}%
^{+}-\tilde{B}_{123}^{+}-\tilde{B}_{124}^{+})\frac{{}}{{}}\right.
\]%
\[
+\frac{1}{24}\left(  B_{144}^{+}B_{322}^{+}+B_{244}^{+}B_{311}^{+}%
-B_{344}^{+}B_{211}^{+}\right)
\]%
\[
+\frac{1}{12}\left(  B_{144}^{+}\tilde{B}_{324}^{+}+B_{244}^{+}\tilde
{B}_{314}^{+}-B_{344}^{+}\tilde{B}_{214}^{+}\right)
\]%
\begin{equation}
+\left.  \frac{1}{12}\left(  B_{144}^{-}B_{324}^{-}+B_{244}^{-}B_{314}%
^{-}-B_{344}^{-}B_{214}^{-}-B_{144}^{-}B_{422}^{-}\right)  \right]
+(2\leftrightarrow3)+(1\leftrightarrow3).
\end{equation}

\section{C-odd exchange}
Evolution equation (\ref{evolutionEq2}) for the odderon exchange reads%
\[
\frac{\partial B_{123}^{-}}{\partial\eta}=\frac{\alpha_{s}3}{4\pi^{2}}\int
d\vec{z}_{4}\frac{\vec{z}_{12}^{\,\,2}}{\vec{z}_{14}^{\,\,2}\vec{z}%
_{42}^{\,\,2}}\left[  B_{423}^{-}+B_{143}^{-}-B_{123}^{-}\frac{{}}{{}}\right.
\]%
\[
-B_{124}^{-}-B_{443}^{-}+B_{424}^{-}+B_{144}^{-}+\frac{1}{12}\left(
B_{144}^{+}B_{324}^{-}+B_{244}^{+}B_{314}^{-}-B_{344}^{+}B_{214}^{-}\right)
\]%
\begin{equation}
+\left.  \frac{1}{12}\left(  B_{144}^{-}B_{324}^{+}+B_{244}^{-}B_{314}%
^{+}-B_{344}^{-}B_{214}^{+}\right)  \right]  +(2\leftrightarrow
3)+(1\leftrightarrow3).
\end{equation}
Here, as in the pomeron case one can express the C-even baryon Green functions
through 2- and 4- reggeon pomerons%
\[
\frac{\partial B_{123}^{-}}{\partial\eta}=\frac{\alpha_{s}3}{4\pi^{2}}\int
d\vec{z}_{4}\frac{\vec{z}_{12}^{\,\,2}}{\vec{z}_{14}^{\,\,2}\vec{z}%
_{42}^{\,\,2}}\left[  B_{423}^{-}+B_{143}^{-}-B_{123}^{-}\frac{{}}{{}}\right.
\]%
\[
-B_{124}^{-}-B_{443}^{-}+B_{424}^{-}+B_{144}^{-}+\frac{1}{12}\left(
B_{144}^{+}B_{324}^{-}+B_{244}^{+}B_{314}^{-}-B_{344}^{+}B_{214}^{-}\right)
\]%
\[
+\frac{1}{24}\left(  B_{144}^{-}(B_{322}^{+}+B_{224}^{+}+B_{334}%
^{+})+B_{244}^{-}(B_{311}^{+}+B_{344}^{+}+B_{114}^{+})\right)
\]\[
-\frac{1}{24}B_{344}^{-}(B_{211}%
^{+}+B_{114}^{+}+B_{224}^{+})
\]%
\begin{equation}
+\left.  \frac{1}{12}\left(  B_{144}^{-}\tilde{B}_{324}^{+}+B_{244}%
^{-}\tilde{B}_{314}^{+}-B_{344}^{-}\tilde{B}_{214}^{+}\right)  \right]
+(2\leftrightarrow3)+(1\leftrightarrow3).
\end{equation}

\section{Conclusion}

In this paper evolution equations for the C-odd and
C-even Green functions for the 3 quark Wilson loop are
presented in the leading logarithm approximation. The method of high energy operator expansion developed in
\cite{Balitsky},\cite{Balitsky:2010jf} was used and the result coincides with the dipole BK equation
if two of the quarks have the same transverse coordinates. As in
\cite{Bartels:2007aa} the C-even Green function obtained here consists of 3
2-reggeon BFKL pomerons and one 4-reggeon term, while the C-odd term consists
of only one 3-reggeon odderon. All the nonlinear terms describing the
interaction of these three Green functions are presented. The 4-reggeon C-even Green function starts from the 4 gluon exchange and therefore it gives a next-to-leading contribution to amplitudes. 

In large $N_c$ limit one can treat the matrix element of a product of colorless gauge-invariant operators as a product of the corresponding matrix elements. Going to this limit makes the BK equation closed. The derivation given here can be straightforwardly generalized to the colorless baryon-like state in SU($N_c$) for $N_c>3$. However, the evolution equation will change not only in the coefficients, but in the form. Therefore taking large $N_c$ limit in such a way will not lead to the physical SU(3) baryon-like Green function. Nevertheless, one may try fixing the form of the equation and formally imposing the large $N_c$ limit on it to get the closed expression. In this case one should consider the Green function for the 3-quark Wilson loop with products of the SU(3) Levi-Civita tensors expanded in terms of Kronecker deltas.

\acknowledgments
We would like to thank all the participants of the theoretical seminar at Budker Institute for critical discussion of this work. We thank prof. V. S. Fadin, M. G. Kozlov and A.V. Reznichenko for helpful discussions. A. V. G. is grateful to prof. L. Szymanowski for bringing this topic to his attention. R. E. G. is grateful to the Dynasty foundation for financial support. The study was supported by the Ministry of education of the Russian Federation projects 14.B37.21.1181 and 8408 and by the Russian Fund for Basic Research grants 12-02-31086, 10-02-01238 and 12-02-33140.

\end{document}